

\magnification = \magstep1
\hsize = 16 truecm
\parskip = 0pt
\normalbaselineskip = 19pt plus 0.2pt minus 0.1pt
\baselineskip = \normalbaselineskip
%

\font\brm=cmr10 scaled \magstep1
\font\bbf=cmbx10 scaled \magstep1
\font\bit=cmti10 scaled \magstep1



%
\def\Bigbreak{\par \ifdim\lastskip < \bigskipamount \removelastskip \fi
                   \penalty-300 \vskip 10mm plus 5mm minus 2mm}
%
\newcount\eqnum
\newcount\tempeq
\def\cleareqnum{\global\eqnum = 0}
\def\eqname{(\the\chnum.\the\eqnum)}
\def\neweq{\global\advance\eqnum by 1 \eqno\eqname}
\def\neweqalign{\global\advance\eqnum by 1 &\eqname}
\def\releq#1{\global\tempeq=\eqnum \advance\tempeq by #1
             (\the\chnum.\the\tempeq)}

\cleareqnum
%
\newcount\chnum
\def\clearchnum{\global\chnum = 0}

\def\newchapt#1{\Bigbreak \global\advance\chnum by 1
                \cleareqnum
                \centerline{\bbf\the\chnum.{ }#1}
                \nobreak\vskip 5mm plus 2mm minus 1mm}
\clearchnum
%
\newcount\notenumber
\def\clearnotenumber{\notenumber=0}
\def\note{\advance\notenumber by1 \footnote{$^{\the\notenumber}$}}
\clearnotenumber
%
\newbox\Eqa
\newbox\Eqb
\newbox\Eqc
\newbox\Eqd
\newbox\Eqe
\newbox\Eqf
\newbox\Eqg
\newbox\Eqh
\newbox\Eqi
\def\storeeq#1{\setbox #1=\hbox{\eqname}}

\pageno=0
\footline={\ifnum\pageno=0\strut\hfil\else\hfil\tenrm\folio\hfil\strut\fi}
\def\H{{\cal H}}
\def\D{{\cal D}}
\def\L{{\cal L}}
\def\R{{\bf R}}
\def\C{{\bf C}}
\def\F{{\cal F}}
\def\P{{\cal P}}

\def\Z{{\bf Z}}
\def\b{b(f)}
\def\bc{b^\ast (f)}
\def\ba{b^\sharp (f)}
\def\a{a(f)}
\def\ac{a^\ast (f)}
\def\acg{a^\ast (g)}
\def\aa{a^\sharp (f)}

%


\rightline {IFUP-TH 51/93}
\vskip 1.5 truecm
\centerline {\bbf Fock Representations of Quantum Fields}
\smallskip
\centerline {\bbf with Generalized Statistics}
\vskip 1.3 truecm
\centerline {\brm A. Liguori}
\medskip
\centerline {\bit Dipartimento di Fisica dell'Universit\`a di Pisa}
\medskip
\bigskip
\centerline {\brm M. Mintchev}
\medskip
\centerline {\bit Istituto Nazionale di Fisica Nucleare, Sezione di Pisa}
\centerline {\bit Dipartimento di Fisica dell'Universit\`a di Pisa}
\vskip 3.5 truecm
\centerline {\bit Abstract}
\bigskip

We develop a rigorous framework for constructing Fock representations of
quantum fields obeying generalized statistics associated with certain
solutions of the spectral quantum Yang-Baxter equation. The main features
of these representations are investigated. Various aspects of the underlying
mathematical structure are illustrated by means of explicit examples.
\vskip 2.5 truecm
\centerline {October 1993}
\vfill \eject


\newchapt {Introduction}

The present paper concerns a generalization of the concept of
statistic in quantum field theory. We investigate the
Fock realization of $N$-component fields whose annihilation parts obey
the relation
$$
a_\alpha (x_1)a_\beta (x_2) =
R_{\beta \alpha }^{\delta \gamma }(x_2,x_1)\,
a_\gamma (x_2)a_\delta (x_1) \quad .
\neweq
$$
Here $x_1, x_2 \in \R^s$ and the {\it exchange factor} $R$ is a
$N^2\times N^2$ matrix function on $\R^s\times \R^s$, which
satisfies certain consistency conditions to be specified in what follows.
Our main task below is to construct a Fock representation $\F_R$ of the
algebra (1.1). We also establish the
basic properties of the fields $a_\alpha (x)$ as operators in $\F_R$
and derive the relative correlation functions.
Finally, some aspects of the time evolution in the Fock space $\F_R$
are investigated.

The statistic of quantum fields is usually
associated with the irreducible representations of the permutation group
which lead to bosons (more generally para-bosons) and fermions
(para-fermions). In the last two decades
it has been discovered however that alternative, not
permutation group statistics [1-5] appear actually in many different areas of
quantum theory. An outstanding example is conformal field theory;
the basic building blocks of conformal invariant models
in 1+1 space-time dimensions obey statistics corresponding in general to
irreducible representations of the braid instead of the permutation group
[6]. Charged sectors of 2+1 dimensional
gauge theories [7] also give rise to braid statistics [8]. The excitations
associated with Abelian and non-Abelian representations of the
braid group are called anyons and plektons respectively.
It is generally believed
[9] that anyons occur as quasi-particles in the fractional quantum Hall
effect and may be relevant for explaining some features of high
$T_c$-superconductors [10]. Other striking physical phenomena as
fractional charge [11] and fractional spin [4] are also deeply related to
generalized statistics. For all these reasons the subject has been
extensively studied in the last few years (see [12]
and references therein). The methods which have been most frequently applied
are the mean-field approach [13] and the non-relativistic
field theory formulation of the $N$-body quantum-mechanical anyon
problem [14].

One of the main goals of this work is to show that the Fock representation of
the algebra (1.1) provides a convenient basis and a unifying framework
for the investigation of statistics in quantum field
theory. Indeed, varying $R$ among the admissible
exchange factors, one gets a rich family of quantum fields.
The $x$-independent $R$-matrices lead to permutation group statistics
whereas piecewise-constant $R$-factors give rise to braid
statistics. But the family in consideration is even larger because it
involves also fields corresponding to $R$-matrices with
more complicated $x$-dependence. From the mathematical point of view
we find it instructive to study the whole family
in general. Concerning physics, one should mention that
deviations from the Bose-Fermi alternative are not expected for fundamental
elementary particles [15]. There exist several indications however, that
generalized statistics are relevant for describing
collective excitations in solid state physics [9,10].
Products of ``order" and ``disorder" variables [16],
which control the phase structure
in quantum field theory and statistical mechanics, are also expected to
have exotic statistics [17].

We end the introduction by recalling those universal structures
of any Fock representation (see for example section X.7 of [18]),
which are used below. This allows both to fix the notation
and to explain our strategy for constructing the Fock representation
$\F_R$. Consider a separable
Hilbert space $\{\, \H \, ,\, (\cdot \, ,\, \cdot )\, \}$ and its
$n$-fold tensor power $\H^n = \H^{\otimes n}$ which in physical
terms is the $n$-particle space. The direct sum
$$
\F (\H ) = \bigoplus_{n=0}^\infty \, \H^n \quad , \neweq
$$
where $\H^0 = \C^1$ is called the Fock space over $\H$.
The elements of $\F (\H )$ can be represented by sequences
$\{ \varphi = (\varphi^{(0)}, \varphi^{(1)},...,\varphi^{(n)},... ) \,
\, : \, \, \varphi^{(n)} \in \H^n \}$ and
the finite particle subspace $\F^0 (\H )\subset \F (\H )$ is defined as
follows: $\varphi \in \F^0 (\H )$ if and only if
$\varphi^{(n)} = 0$ for $n$ large enough. By
construction $\F^0 (\H )$ is dense in $\F (\H )$.

We turn now to the definition of annihilation and creation operators on
$\F^0 (\H )$. Let us denote by $\D^n$ the set of decomposable vectors
$$
\D^n = \{\, f_1\otimes \cdots \otimes f_n \, :\, f_i \in \H \, \}
\subset \H^n \quad , \quad \quad \D^0 = \C^1 \quad .
$$
Recall that $\D^n$ is a total set, i. e.
the set $\L (\D^n )$ of finite linear combinations of
elements of $\D^n $ is dense in $\H^n$. Following [18], for each
$f\in \H$ we introduce the operators
$$
\eqalignno{\b \, :\, \D^n &
\longrightarrow \D^{n-1}\quad , \quad \quad n \geq 1
\quad , \cr
\bc \, :\, \D^n &\longrightarrow \D^{n+1}\quad , \quad \quad n\geq 0
\quad , \cr}
$$
defined by
$$
\b \, f_1\otimes \cdots \otimes f_n =
\sqrt{n}\, (f,f_1)\, f_2\otimes \cdots \otimes f_n
\quad , \neweq
$$
$$
\bc \, f_1\otimes \cdots \otimes f_n =
\sqrt{n+1}\, f\otimes f_1\otimes \cdots \otimes f_n \quad . \neweq
$$
In addition, we set $\b \H^0 = 0$
and extend by linearity both $\b$ and $\bc$ to $\L (\D^n )$.
Now, one can easily prove [18] that for any $\varphi \in \L (\D^n )$ and
$\psi \in \L (\D^{n+1} )$ one has:
$$
\Vert \b \varphi \Vert \leq \sqrt{n}\, \Vert f \Vert \, \Vert \varphi \Vert
\quad , \quad \quad
\Vert \bc \varphi \Vert \leq \sqrt{n+1}\, \Vert f \Vert \, \Vert \varphi \Vert
\quad , \neweq
$$
$$
(\psi \, , \, \bc \varphi ) = (\b \psi \, , \, \varphi ) \quad . \neweq
$$
\medskip
In what follows $b^\sharp$ denotes either $b$ or $b^\ast $.
{}From the estimates (1.5) one has that
$\ba $ can be extended by continuity
to $\H^n$ and finally, by linearity, to $\F^0 (\H )$.
The extensions obviously obey the relation (1.6) for any
$\varphi, \psi \in \F^0 (\H )$.

For describing bosons and fermions
in this context one introduces the subspaces $\H_+^n$
and $\H_-^n$ of $\H^n$, which are the $n$-fold symmetric and totally
anti-symmetric tensor powers of $\H$ respectively. Then,
the associated bosonic and fermionic Fock spaces are
$$
\F_\pm (\H ) = \bigoplus_{n=0}^\infty \, \H_\pm ^n \quad . \neweq
$$
Denoting by $P_\pm $ the orthogonal projectors on $\F_\pm (\H )$, the
bosonic and fermionic creation and annihilation operators are defined by
$$
a_\pm^\sharp (f) = P_\pm \, \ba \, P_\pm \quad . \neweq
$$

At this point we have a sufficient background for formulating
our approach to the Fock realization of the algebra (1.1).
The main steps are essentially two:

\item {(i)} with any admissible exchange factor $R(x_1,x_2)$ we
associate a distinguished subspace $\F_R (\H )\subset \F (\H )$;

\item {(ii)} we generalize eq.(1.8) by replacing $P_\pm$ with the
orthogonal projection $P_R$ on $\F_R (\H )$.

\noindent The key points of this procedure
are discussed in full details
in the next two sections. Section 4 concerns the description of time
evolution in our framework. In section 5 we show that the
Leinaas-Myrheim anyons represent just a particular example of
the general scheme developed in sections 2-5. In section 6 we
establish the generalization to multicomponent fields.
Finally, section 7 is devoted to our conclusions.

\newchapt {Exchange Factors and Relative Fock Spaces}

For simplicity we start by considering eq.(1.1) for fields with
a single component ($N=1$). We take as one-particle space $\H$
the complex Hilbert space $L^2(\R^s, d^sx )$, the generalization to an
arbitrary $L^2({\bf X}, d\mu )$ being straightforward.
With these assumptions,
the exchange factor $R$ is in general a complex-valued function on
$\R^s \times \R^s$. We require $R$ to be measurable and to satisfy
$$
R(x_1,x_2)\, R(x_2,x_1) = 1 \quad , \neweq
$$
$$
{\overline R}(x_1,x_2) = R(x_2,x_1)  \quad , \neweq
$$
where the bar stands for complex conjugation. Eq.(2.1) guaranties the
consistency of eq.(1.1) under the interchange of $x_1$ and $x_2$.
The meaning of (2.2) will be clarified a few lines below.

Combining eqs.(2.1,2) one gets the parametrization
$$
R(x_1,x_2) = \exp [ i r(x_1,x_2) ] \quad , \neweq
$$
where $r$ is real-valued and obeys
$$
r(x_1,x_2) + r(x_2,x_1)  = 2 \pi k_r \quad , \quad \quad k_r \in \Z \quad .
\neweq
$$
We call $R$
{\it factorizable} if there exists a real-valued function $p$ on
$\R^s$ and $k_r \in \Z$, such that
$$
r(x_1,x_2) = p(x_1) - p(x_2) + \pi k_r \quad . \neweq
$$
Notice that in general $R(x_1,x_2)^2 \not= 1$. For $N=1$ the only
possible {\it constant} exchange factors are $\pm 1$, i. e. generalized
statistics require $x$-dependent $R$.

Fixing any admissible exchange factor, we introduce for $n\geq 2$
the operators\break $\{\, S_i\, :\, i=1,...,n-1\, \}$ acting in
$\H^n$ according to
$$
\left [S_i\varphi \right ](x_1,...,x_i,x_{i+1},...,x_n ) =
R(x_i,x_{i+1})\varphi (x_1,...,x_{i+1},x_i,...,x_n ) \quad .\neweq
$$
As a simple consequence of the defining properties of $R$, one has
\medskip

{\bf Proposition 1}: $\{\, S_i\, :\, i=1,...,n-1\, \}$ {\it are bounded}
($\Vert S_i \Vert = 1$) {\it Hermitian operators on} $\H^n$
{\it satisfying}:
$$
S_i\, S_j = S_j\, S_i \quad , \quad \quad |i-j| \geq 2 \quad ,
$$
$$
S_i\, S_{i+1}\, S_i = S_{i+1}\, S_i\, S_{i+1} \quad , \quad \quad
S_i^2 = 1 \quad .
$$
\medskip
\noindent We observe that (2.2) ensures the hermiticity of $S_i$,
which is actually the main motivation for this condition on $R$.

Let ${\cal P}_n$ be the permutation group of
$n$ elements, whose generators (elementary permutations) are denoted by
$\{\, \sigma_i \, :\, i = 1,...,n-1\, \}$. The above proposition has the
following simple
\medskip

{\bf Corollary}: {\it The mapping}
$$
S\, :\, \sigma_i \longmapsto S_i  \neweq
$$
{\it provides a representation of} ${\cal P}_n$ {\it in} $\H^n$ {\it and}
$$
P_R^{(n)} = {1\over n!}\, \sum_{\sigma \in {\cal P}_n} \, S(\sigma )
\quad , \quad \quad n\geq 2 \neweq
$$
{\it is an orthogonal projection operator}.
\medskip

Setting $P_R^{(0)} = {\bf 1}$ and $P_R^{(1)} = {\bf 1}$, we
define the $n$-particle space relative to $R$ by
$$
\H_R^n = P_R^{(n)}\, \H^n \quad . \neweq
$$
The set $\{\, P_R^{(n)}\, :\, n = 0,1,...\, \}$
determines a projection operator $P_R$ on $\F (\H )$, given by
$$
\left [P_R \varphi \right ]^{(n)} = P_R^{(n)}\, \varphi^{(n)}
\quad . \neweq
$$
It is worth mentioning that $P_R$ for $R=\pm 1$ coincides precisely with
$P_\pm $. The $R$-subspace $\F_R (\H )\subset \F (\H )$ we are looking for
is defined by
$$
\F_R (\H ) = P_R\, \F (\H ) = \bigoplus_{n=0}^\infty \, \H_R ^n
\quad . \neweq
$$
The associated finite particle subspace reads
$\F^0_R (\H ) = P_R\, \F^0 (\H )$ and is dense in $\F_R (\H )$.

Consider now the operator $C$, defined on $\H^n$ by
$$
(C\varphi )(x_1,...,x_n) = {\overline \varphi}(x_n,...,x_1) \quad .
\neweq
$$
One immediately verifies that $C$ is antilinear, norm-preserving and
satisfies $C^2 = {\bf 1}$.
Therefore $C$ represents a {\it conjugation} on $\H^n$, which automatically
extends to a conjugation on $\F (\H )$.
{}From eqs.(2.2,6) it follows that $C\, S_i\, C = S_{n-i}$. Consequently
$C\, \F_R (\H ) \subset \F_R (\H )$ , i. e. $C$ provides a conjugation
in any $R$-subspace $\F_R (\H )$.

We conclude this section by establishing the relationship between any couple
of Fock spaces $\F_{R_1} (\H )$ and $\F_{R_2} (\H )$ relative to two
different exchange factors $R_1$ and $R_2$. Clearly, being infinite
dimensional separable Hilbert spaces,
$\F_{R_1} (\H )$ and $\F_{R_2} (\H )$
are isomorphic. Among others, there exist however some natural isomorphisms
which play a distinguished role and can be constructed as follows. Let us
consider the functional equation
$$
T_\pm (x_1,x_2)\, R(x_1,x_2) = \pm \, T_\pm (x_2,x_1)  \neweq
$$
with the supplementary condition
$$
{\overline T_\pm } (x_1,x_2)\, T_\pm (x_1,x_2) = 1 \quad . \neweq
$$
Eqs.(2.13,14) have several solutions; recalling eqs.(2.3,4),
the most evident ones are
$$
\exp \left [ -{i\over 2}r(x_1,x_2) \right ] =
\cases{T_+ (x_1,x_2)   & if $k_r \in 2\Z $ ;  \cr
       T_- (x_1,x_2)   & if $k_r \in 2\Z + 1 $ . \cr }
\neweq
$$


For any solution $T_\pm (x_1,x_2)$ of (2.13,14) and
for any $\varphi \in \H^n$ we define the operators
$$
\left [U_\pm ^{(n)}(R)\varphi \right ](x_1,...,x_n) =
\left [ \prod_{i,j=1 \atop i<j}^n T_\pm (x_i,x_j) \right ]
\varphi (x_1,...,x_n) \quad , \quad \quad n \geq 2 \quad , \neweq
$$
setting also $U_\pm ^{(0)}(R) = {\bf 1}$ and
$U_\pm ^{(1)}(R) = {\bf 1}$.
Due to eq.(2.14), $U^{(n)}_\pm (R)$ are unitary operators on $\H^n$,
which therefore extend to unitary operators $U_\pm (R)$ on $\F (\H )$.
Using (2.13), one immediately verifies that
$$
U_\pm (R) \, :\, \F_R(\H ) \longrightarrow \F_\pm (\H ) \quad . \neweq
$$
Consequently the compositions
$$
U_\pm (R_1,R_2) = U_\pm (R_1)^{-1}\, U_\pm (R_2)  \neweq
$$
provide two natural isomorphisms
between $\F_{R_2} (\H )$ and $\F_{R_1} (\H )$. Notice that the operators
$C_\pm = U_\pm (R)\, C\, U_\pm (R)^{-1}$ are
conjugations on $\F_\pm (\H )$; one has
$$
\left [ C_\pm \varphi \right ](x_1,...,x_n) =
\left [ \prod_{i,j =1 \atop i\neq j}^n T_\pm (x_i,x_j) \right ]
{\overline \varphi }(x_n,...,x_1) \quad , \neweq
$$
which in general differ from $C$.

Summarising, for any admissible exchange factor $R$ we have explicitly
constructed an $R$-subspace $\F_R (\H )\subset \F (\H )$. Any
$\F_R (\H )$ is equipped with a conjugation and we have
established some isomorphisms between pairs of such Fock spaces.
Our next step will be to define
creation and annihilation operators on $\F_R (\H )$.

\newchapt {Creation and Annihilation Operators}

In analogy with eq.(1.8), we introduce the creation and annihilation
operators
$$
\aa = P_R\, \ba \, P_R \quad . \neweq
$$
The estimates (1.5) imply that $\aa$ are densely defined
(with domain $\F^0 (\H )$) linear operators, which satisfy
$$
(\psi \, , \, \ac \varphi ) = (\a \psi \, , \, \varphi ) \quad . \neweq
$$
Therefore $\aa$ are closable. Since
$\F^0_R (\H )$ is invariant under $\aa $, we shall concentrate
below on the restrictions of $\a $ and $\ac $ to $\F^0_R (\H )$.
Their action is given by
$$
\left [\a \varphi \right ]^{(n)}(x_1,...,x_n) =
\sqrt{n+1} \int d^sx\, {\overline f}(x)\varphi^{(n+1)} (x, x_1,...,x_n)
\quad , \neweq
$$
$$
\left [\ac \varphi \right ]^{(n)}(x_1,...,x_n) =
{1\over \sqrt{n}}
f(x_1)\varphi^{(n-1)} (x_2,...,x_n) +
$$
$$
{1\over \sqrt{n}} \sum_{k=2}^{n}\, R(x_{k-1},x_k)...R(x_1,x_k)
f(x_k)\varphi^{(n-1)} (x_1,...,{\widehat x_k},...,x_n)
\quad , \neweq
$$
where $\varphi \in \F^0_R (\H )$ and
$\widehat x$ indicates that the argument $x$ must be omitted.
For deriving the commutation properties of
$\aa$ it is convenient to introduce the operator-valued distributions
$a^\sharp (x)$ defined by
$$
\a = \int d^sx\, {\overline f}(x)\, a (x) \quad , \quad \quad
\ac = \int d^sx\, f(x)\, a^\ast (x) \quad .
$$
Then from eqs.(3.3,4) one gets
$$
\left [a(x) \varphi \right ]^{(n)}(x_1,...,x_n) =
\sqrt{n+1}\, \varphi^{(n+1)} (x, x_1,...,x_n)
\quad , \neweq
$$
$$
\left [a^\ast (x) \varphi \right ]^{(n)}(x_1,...,x_n) =
{1\over \sqrt{n}}
\delta (x-x_1)\varphi^{(n-1)} (x_2,...,x_n) +
$$
$$
{1\over \sqrt{n}} \sum_{k=2}^{n}\, R(x_1,x_k)...R(x_{k-1},x_k)
\delta (x-x_k)\varphi^{(n-1)} (x_1,...,{\widehat x_k},...,x_n)
\quad . \neweq
$$
Precisely as for bosons and fermions, $a(x)$ is a densely defined
operator in $\F_R (\H )$ whereas $a^\ast (x)$
makes sense only as a densely defined
quadratic form in $\F_R (\H ) \times \F_R (\H )$. Using eqs.(3.5,6),
one easily checks that $a^\sharp (x)$ satisfy the exchange relations
$$
a(x_1)\, a(x_2) - R(x_2,x_1)\, a(x_2)\, a(x_1) = 0
\quad , \neweq
$$
$$
a^\ast (x_1)\, a^\ast (x_2) - R(x_2,x_1)\,
a^\ast (x_2)\, a^\ast (x_1) = 0 \quad , \neweq
$$
$$
a(x_1)\, a^\ast (x_2) - R(x_1,x_2)\, a^\ast (x_2)\, a(x_1) =
\delta (x_1-x_2) \quad . \neweq
$$

Now we are going to establish some other basic features of the fields
$\aa $. It is well known that $a_-^\sharp (f)$ are bounded operators,
which is not the case of $a_+^\sharp (f)$. The following proposition
generalizes this statement.
\medskip

{\bf Proposition 2}: $\aa$ {\it are bounded operators of norm}
$$
\Vert \aa \Vert \leq \Vert f \Vert \quad , \neweq
$$
{\it if and only if}
$$
\int d^sy \int d^sz\, {\overline g}(y)\, R(y,z)\, g(z) \leq 0 \neweq
$$
{\it for any} $g\in B_0 (\R^s )$ - {\it the space of bounded functions
with compact support}.
\medskip

{\it Proof}: Because of eq.(3.2), it is sufficient to consider $\ac $.
Using eqs.(3.3,4), one finds the equality
$$
\Vert \acg \varphi \Vert^2 = \Vert g \Vert^2 \Vert \varphi \Vert^2 +
$$
$$
n\int d^sx_1 \dots \int d^sx_{n-1} \int d^sy \int d^sz
{\overline g}(z)\, R(y,z) g(y) \varphi (z,x_1,...,x_{n-1})
{\overline \varphi}(y,x_1,...,x_{n-1}) \,  \, , \neweq
$$
valid for any $g\in B_0 (\R^s )$ and $\varphi \in \H^n_R$.

Assume first that $\ac$ is bounded and satisfies (3.10). Taking
$\varphi \in \H$ in eq.(3.12), one gets
$$
\Vert g \Vert^2 \Vert \varphi \Vert^2 +
\int d^sy \int d^sz
{\overline g}(z)\, R(y,z) g(y) \varphi (z)
{\overline \varphi}(y) =
\Vert \acg \varphi \Vert^2 \leq
\Vert \acg \Vert^2 \Vert \varphi \Vert^2
\quad , \neweq
$$
which in view of eq.(3.10) leads to
$$
\int d^sy \int d^sz {\overline \varphi}(y) g(y) \, R(y,z)\,
\varphi (z) {\overline g}(z)  \leq 0 \neweq
$$
for arbitrary $g\in B_0 (\R^s )$ and $\varphi \in \H$.
This proves (3.11).

Suppose now that (3.11) holds and consider
$$
\chi (x_1,...,x_{n-1}) =
\int d^sy \int d^sz
{\overline g}(z)\, R(y,z) g(y) \varphi (z,x_1,...,x_{n-1})
{\overline \varphi}(y,x_1,...,x_{n-1}) \, . \neweq
$$
For any $g\in B_0 (\R^s )$ one has that
$\chi \in L^1(\R^{s(n-1)})$. Moreover, from
(3.11) it follows that $\chi (x_1,...,x_{n-1}) \leq 0$ almost everywhere
in $\R^{s(n-1)}$. Therefore
$$
\int d^sx_1 \dots \int d^sx_{n-1} \chi (x_1,...,x_{n-1})
\leq 0 \quad ,
$$
which, combined with eq.(3.12) gives
$$
\Vert \acg \varphi \Vert \leq \Vert g \Vert \, \Vert \varphi \Vert \quad .
\neweq
$$
This estimate holds obviously also for any $\varphi \in \F^0_R (\H )$ and
implies (3.10) because $B_0 (\R^s )$ and $\F^0_R (\H )$ are dense
in $\H$ and $\F_R (\H )$ respectively.
\medskip

We observe that any factorizable exchange factor (see eq.(2.5)) with odd
values of $k_r$ obeys (3.11) and consequently leads to bounded
creation and annihilation operators.
In general $\aa $ are not bounded, but the technical difficulties
stemming from this fact can be avoided (at least partially) by
considering bounded functions of $\aa $. Following for instance
the standard treatment [18] of the boson field $a_+^\sharp (f)$,
one can introduce the Segal-type operator
$$
\Phi (f) = {1\over \sqrt{2}} \left [\a + \ac \right ] \neweq
$$
and prove exactly in the same way
\medskip

{\bf Proposition 3}: {\it The operator} $\Phi (f)$
{\it is essentially self-adjoint on} $\F^0_R (\H )$.
\medskip

Therefore, inspite of the fact that in general
$\Phi (f)$ is unbounded, the operator $\exp \left [i\Phi (f)\right ] $
built with the self-adjoint closure of (3.17), is a unitary operator
on $\F_R (\H )$. Notice that the Segal
field (3.17) is not a linear functional in $f$ since $\a $ is
antilinear.

We now turn to the construction of the correlation functions of the
fields $\aa $. Denoting the vacuum state by
$\Omega = (1,0,...,0,...)$ and
using that ${\cal D}^n$ is a total set in $\H^n$, one can show that
$\{ a^\ast (f_1) \cdots a^\ast (f_n)\Omega \, : \, f_i \in \H \}$
is a total set in $\H_R^n$. Therefore, all non-trivial correlation
functions (in the form of distributions) are
$$
w_n (x_1,...,x_n;y_1,...,y_n) =
\bigl ( a^\ast (x_1) \cdots a^\ast (x_n)\Omega \, , \,
a^\ast (y_1) \cdots a^\ast (y_n)\Omega \bigr ) \, \, , \quad
n\geq 0 \quad . \neweq
$$
Applying eq.(3.2,9) one derives the recursive relation
$$
w_n (x_1,...,x_n;y_1,...,y_n) = \delta (x_1-y_1)
w_{n-1} (x_2,...,x_n;y_2,...,y_n)  +
$$
$$
\sum_{k=2}^n R(x_1,y_1)...R(x_1,y_{k-1})\, \delta (x_1-y_k)\,
w_{n-1} (x_2,...,x_n;y_1,...,{\widehat y}_k,...,y_n) \quad , \neweq
$$
which permits to compute $w_n$. One finds
$$
w_0 = (\Omega \, ,\, \Omega ) = 1 \quad , \quad \quad
w_1(x_1;y_1) = \delta (x_1-y_1) \quad ,
$$
$$
w_2(x_1,x_2;y_1,y_2) = \delta (x_1-y_1)\, \delta (x_2-y_2) +
R(x_1,y_1)\, \delta (x_1-y_2)\, \delta (x_2-y_1)
$$
and so on. As expected the $R$-dependence shows up for $n\geq 2$.

\newchapt {Time Evolution}

We first recall the notion of {\it second quantization} $d\Gamma (A)$ of an
operator $A$ acting in the one-particle space $\H$. Assume that
$A$ has a dense domain $D\subset \H$. Then the subset
$D(A)= \{\, \varphi \in \F^0 (\H )\, :\, \varphi^{(n)}\in D^{\otimes n}
\, \, \, {\rm for}\, \, \,  n\geq 1 \}$ is dense in $\F (\H )$ and
$d\Gamma (A)$ is defined on $D(A)$ by [18]
$$
\left [d\Gamma (A)\varphi \right ]^{(0)} = 0 \quad , \neweq
$$
$$
\left [d\Gamma (A)\varphi \right ]^{(n)} =
(A\otimes {\bf 1}\otimes \cdots \otimes {\bf 1} +
{\bf 1}\otimes A\otimes \cdots \otimes {\bf 1} + ... +
{\bf 1}\otimes {\bf 1}\otimes \cdots \otimes A )\varphi^{(n)} \, \, .
\neweq
$$
We furthermore introduce the operator
$$
d\Gamma_R (A) = P_R \, d\Gamma (A) \, P_R \quad . \neweq
$$
Provided that
$$
P_R \, D(A) \subset D(A) \quad , \neweq
$$
eq.(4.3) defines an operator in $\F_R (\H )$ with dense domain
$D_R(A) = P_R\, D(A)$. For\break $R=\pm 1$ the condition (4.4)
is automatically satisfied and eq.(4.3) gives rise to the operators
$d\Gamma_\pm (A) = P_\pm d\Gamma (A)P_\pm $ on $D_\pm (A) = P_\pm D(A)$.
These operators are familiar from the Bose- and Fermi-quantization.

Let us consider now the time evolution in $\F_R (\H )$ starting
with the free one-particle hamiltonian
$$
h = -{1\over 2}\Delta \quad , \neweq
$$
defined on the space of $C^\infty $-functions with compact
support $C^\infty_0(\R^s)\subset L^2(\R^s)$.
It is well known that $h$ is essentially self-adjoint.
In this section we concentrate on smooth exchange factors
$R\in C^\infty (\R^s\times \R^s)$. For such factors
$P_R\, D(h)\subset D(h)$ and therefore
$$
H_R = d\Gamma_R(h) = P_R \, d\Gamma (h)\, P_R  \neweq
$$
is a densely defined Hermitian operator in $\F_R (\H )$ with domain
$D_R(h)$. Using that
$P_R$ and $d\Gamma (h)$ commute with the conjugation $C$, which in turn
leaves invariant $D_R(h)$, one concludes
that $H_R$ admits at least one self-adjoint extension [18].
We shall prove now that $H_R$ has a unique self-adjoint extension.
For this purpose we introduce the operators ${\widetilde H}_\pm $
related to $H_R$ by the isospectral transformation
$$
{\widetilde H}_\pm = U_\pm (R)\, H_R\, U_\pm (R)^{-1} \quad . \neweq
$$
${\widetilde H}_\pm $ act in $\F_\pm (\H )$ but, as
shown in what follows, are different and
should be distinguished from the free bosonic and fermionic
hamiltonians $H_\pm = P_\pm \, d\Gamma (h)\, P_\pm $. In order to determine
the domains of ${\widetilde H}_\pm $, one can use the operator identity
$$
U_\pm (R)\, P_R = P_\pm \, U_\pm (R) \quad , \neweq
$$
valid on $\F (\H )$. Applying eq.(4.8), one obtains the chain of equalities
$$
U_\pm (R) D_R(h) =U_\pm (R) P_R D(h)
=  P_\pm U_\pm (R) D(h) = P_\pm D(h)=D_\pm (h) \quad ,
$$
which show that ${\widetilde H}_\pm $ are well defined on
$D_\pm (h)$ - the domains of $H_\pm $.

The $n$-particle hamiltonian following from eq.(4.6) reads
$$
H_R^{(n)}=
- {1\over 2}\, P_R^{(n)}\left (\sum_{k=1}^n \Delta_k \right ) P_R^{(n)}
\quad , \neweq
$$
where $\Delta_k$ operates on the $k^{{\rm th}}$ variable. Therefore
$$
{\widetilde H}_\pm^{(n)} = - {1\over 2}\, U_\pm^{(n)}(R)\,
P_R^{(n)}\left (\sum_{k=1}^n \Delta_k \right ) P_R^{(n)}\,
U_\pm^{(n)}(R)^{-1} \quad ,
$$
which according to eq.(4.8) can be written in the form
$$
{\widetilde H}_\pm^{(n)} = - {1\over 2}\, P_\pm^{(n)}\, U_\pm^{(n)}(R)
\left (\sum_{k=1}^n \Delta_k \right )
U_\pm^{(n)}(R)^{-1}\, P_\pm^{(n)}  \quad . \neweq
$$
This general result holds for $U_\pm^{(n)}(R)$ constructed
(see eq.(2.16)) in terms of any solution $T_\pm$ of eqs.(2.13,14).
For obtaining more explicit expressions for ${\widetilde H}_\pm^{(n)}$
one should fix some $T_\pm $. At this stage
it is convenient to distinguish two cases; we consider
${\widetilde H}_+^{(n)}$ if $k_r \in 2\Z$ and
${\widetilde H}_-^{(n)}$ if $k_r \in 2\Z +1$.
Then one can adopt for $T_\pm$ the simple expression (2.15). Moreover,
without loss of generality one can assume that
$r\in C^\infty (\R^s \times \R^s)$, because $R$ is smooth. With these
choices one easily derives
$$
{\widetilde H}_\pm^{(n)} = -{1\over 2} P_\pm^{(n)}
 \sum_{k=1}^n \biggl[ \nabla_k +
 {i\over 2} \sum_{i,j =1 \atop i<j}^n
(\nabla_k r_{i\,j}) \biggr]^2 P_\pm^{(n)} \quad ,
\neweq
$$
where the short notation
$$
r_{i\, j} = r(x_i,x_j)
$$
has been introduced. The idea now is to move the
projections $P_\pm^{(n)}$ in front of the r.h.s. of eq.(4.11) to the right,
taking into account at the end that $\bigl[P_\pm^{(n)}\bigr]^2 = P_\pm^{(n)}$.
After some algebra one finds
$$
{\widetilde H}_\pm^{(n)}= {\widetilde H}^{(n)} P_\pm^{(n)} \quad , \neweq
$$
where
$$
{\widetilde H}^{(n)}=
 -{1\over 2} \sum_{k=1}^n \Delta_k + V^{(n)} \quad , \neweq
$$
with
$$
V^{(n)} = {1\over 24} \sum_{k=1}^n \left \{ 2\sum_{i=1 \atop i\not= k}^n
(\nabla_k r_{k\,i})^2 + \left [\sum_{i=1 \atop i\not= k}^n
(\nabla_k r_{k\,i})\right ]^2 \right \} \quad . \neweq
$$
Observing that $V^{(n)}\geq 0$ and $V^{(n)}\in L^2 (\R^{ns} )_{{\rm loc}}$,
by Theorem X.28 of reference [18] one concludes that
${\widetilde H}^{(n)}$ is essentially self-adjoint on
$C_0^\infty (\R^{ns})$. Since $P_\pm^{(n)}$ commute with
${\widetilde H}^{(n)}$, the same conclusion holds for
${\widetilde H}_\pm^{(n)}$ on $D_\pm^n (h)=D_\pm (h)\cap \H_\pm^n$.
Then, a standard argument (cf. section VIII.10 of [18]) implies
that ${\widetilde H}_\pm $ are essentially self-adjoint on $D_\pm (h)$,
which combined with eq.(4.7) proves the validity of
\medskip

{\bf Proposition 4}: $H_R$ {\it has a unique self-adjoint extension}.
\medskip

It is useful for physical considerations to analyze the type of
interactions corresponding to the potential $V^{(n)}$. One easily derives
$$
V^{(n)} = {1\over 2} \sum_{i,j=1\atop i\not= j}^n V_{i \, j}
 +{1\over 6} \sum_{i,j,k=1\atop i\not= j, i\not= k, j\not= k }^n
V_{i\, j\, k}\quad . \neweq
$$
Therefore $V^{(n)}$ involves non-trivial two-body
$$
V_{i\, j} = {1\over 8} \bigl(\nabla_i r(x_i,x_j) \bigr)^2 +
{1\over 8} \bigl(\nabla_j r(x_j,x_i) \bigr)^2  \neweq
$$
and three-body interactions
$$
V_{i\, j\, k} = {1\over 12}\bigl( \nabla_i r(x_i,x_j)\bigr)
                           \bigl( \nabla_i r(x_i,x_k)\bigr)
                           + \hbox{ \rm cyclic perm.}
\neweq
$$

Summarising, the unique self-adjoint extension
of $H_R$ defines a time evolution in $\F_R (\H )$.
By means of $U_\pm(R)$ this evolution can be equivalently transferred to
$\F_\pm (\H )$ and in the $n$-particle spaces $\H_\pm^n$
gives rise to the hamiltonian
${\widetilde H}^{(n)}$. The explicit form of ${\widetilde H}^{(n)}$
(see eqs.(4.13,15-17)) shows that
due to the non-trivial exchange properties (encoded in the projection
$P_R^{(n)}$ in eq.(4.9)), the time evolution
corresponding to the free one-particle
hamiltonian (4.5) is in general a
complicated dynamical problem for $n\geq 2$. In this respect the
appearance of three-body interactions is worth stressing.

Along the above lines one can treat also the problem with external
potential, namely
$$
h = -{1\over 2}\Delta + W(x) \quad . \neweq
$$
The one-particle hamiltonian (4.18) gives rise to
$$
{\widetilde H}_\pm^{(n)} = - {1\over 2}\, P_\pm^{(n)}\, U_\pm^{(n)}(R)
\left (\sum_{k=1}^n \Delta_k \right )
U_\pm^{(n)}(R)^{-1}\, P_\pm^{(n)}
+ \sum_{k=1}^n W_k \quad , \neweq
$$
where $W_k = W(x_k)$.

In conclusion we consider as examples two particular exchange
factors which lead to relatively simple hamiltonians.
For factorizable exchange factors (2.5) one obtains from eqs.(4.13,14)
$$
{\widetilde H}^{(n)}=
 -{1\over 2} \sum_{k=1}^n \Delta_k
 +{1\over 24}(n^2-1) \sum_{k=1}^n \bigl(\nabla p(x_k)\bigr)^2  \quad .
\neweq
$$
Therefore, in this case the two- and three-body interactions give rise to
a sort of effective external potential.

An exactly solvable example with non-factorizable exchange factor is the
following one.  Let $\Omega$ be an
antisymmetric real $s\times s$ matrix. Take
$$
r(x_1,x_2) = \Omega_{\mu \,\nu} x_1^\mu x_2^\nu + \pi k_r \quad ,
\quad \quad \mu , \nu = 1,...,s \quad , \quad \quad  k_r \in \Z \quad ,
\neweq
$$
where hereafter the summation over repeated upper and lower indices
is always understood. Inserting (4.21) in (4.16,17) one finds
$$
\eqalign{
V_{i\, j}     & =-{1\over 8} (\Omega )^2_{\mu \,\nu}
                  (x_i^\mu x_i^\nu + x_j^\mu x_j^\nu) \quad ,\cr
V_{i\, j\, k} & =-{1\over 12} (\Omega )^2_{\mu \,\nu }
                  (x_i^\mu x_j^\nu + x_j^\mu x_k^\nu
                  + x_k^\mu x_i^\nu ) \quad . \cr}
\neweq
$$
In this case the three-body potential is actually a superposition of
two-body interactions and ${\widetilde H}^{(n)}$ takes the form:
$$
{\widetilde H}^{(n)}=
 -{1\over 2} \sum_{k=1}^n \Delta_k
 -{1\over 2} \sum_{i,j=1}^n M_{i\,j} (\Omega )^2_{\mu \,\nu }
  x_i^\mu x_j^\nu
\neweq
$$
with
$$
M_{i\,j}=\cases{
{1\over 4}(n-1)             & if $i=j$     ;   \cr
{1\over 12}(n-2)             & if $i\ne j$      \cr}
\qquad
i,j=1,\dots,n .
\neweq
$$
Being symmetric, $M$ can be diagonalized by an
orthogonal matrix; the relative eigenvalues are:
$$
 m_1=m_2=\cdots=m_{n-1}={2n-1\over 12} \quad ,
\quad \quad m_n={n^2-1\over 12}\quad .
\neweq
$$
Consequently, the hamiltonian (4.23) describes $sn$ harmonic
oscillators with the frequencies $|\omega_\mu |\sqrt {m_i}\, $, where
$\omega_\mu $ are the eigenvalues of $\Omega $.

\newchapt {The Leinaas-Myrheim Anyon}

We already mentioned in the introduction that braid statistics are
incorporated in the above general framework as particular cases. As an
example we consider in this section the Leinaas-Myrheim (L-M) anyon
field. In order to make contact with the L-M approach [2], we set
$s=2$ and consider the relation between two-anyon\break
L-M wave functions $\psi (x_1,x_2)$ and two-boson wave functions
${\widetilde \psi } (x_1,x_2)\in \H_+^2$.
Neglect for a moment the center of mass coordinates and introduce in the
relative space polar coordinates $(\varrho , \phi )$. Then,
according to reference [2], one has
$$
{\widetilde \psi }(\varrho ,\phi)=\exp(-i\vartheta\phi)\,
\psi(\varrho ,\phi) \quad , \neweq
$$
where $\vartheta $ is called {\it statistical} parameter.

Let us denote by ${\rm arg}(x;u)\in [-\pi , \pi )$ the oriented angle
between $x$ and an arbitrary but fixed unit vector $u$.
Usually one takes $u = (1,0)$.
Restoring the coordinates $x_1$ and $x_2$, eq.(5.1) reads
$$
{\widetilde \psi } (x_1,x_2) = T_+(x_1,x_2)\, \psi (x_1,x_2)  \neweq
$$
with
$$
T_+(x_1,x_2) = \exp[- i \vartheta \, {\rm arg}(x_1-x_2;u)] \quad . \neweq
$$
Now, using eq.(2.13), one obtains for the exchange factor of the L-M anyon
field
$$
R(x_1,x_2) = T_+(x_2,x_1)\, T_+(x_1,x_2)^{-1} =
\exp \Bigl(- i \pi \, \vartheta \,
{\rm sgn} \left [(x_1-x_2)^\mu {\widetilde u}_\mu \right ]\Bigr)
\quad , \neweq
$$
where ``sgn" is the sign function and
${\widetilde u}_\mu = \varepsilon_{\mu \nu}u^\nu $ is the vector dual
to $u$. In deriving eq.(5.4) we have used the identity
$$
{\rm arg}(x ; u) - {\rm arg}(-x ; u) =
- \pi {\rm sgn} \left (x^\mu {\widetilde u}_\mu \right )
\quad . \neweq
$$
The expression (5.4) defines a piecewise-constant
admissible exchange factor which
is already familiar [19]; in $2+1$ dimensional gauge
theories it describes the exchange properties of charged fields localized
on strings. So, one can apply
our general procedure and reconstruct the L-M anyon fields
$a(f;u)$ and $a^\ast (f;u)$ as operators
acting in the Fock space $\F_R (L^2 (\R^2 ))$. In particular,
using eq.(3.19) one obtains in explicit form all equal-time
anyon correlation functions.

Let us consider now the time evolution of L-M anyons associated with the
one-particle hamiltonian (4.18). Strictly speaking, one can not apply
directly the results of the previous section, because $R$ is discontinuous
on
$$
\gamma_{12} = \{x_1,x_2 \in \R^2 \, :\,
(x_1-x_2)^\mu {\widetilde u}_\mu = 0 \} \quad . \neweq
$$
In order to avoid this difficulty we introduce the space
${\widetilde D}_+^n$ of $C_0^\infty$-functions in $\H_+^n$,
which vanish with all derivatives on the union of
$\gamma_{ij}$ with $1\leq i < j \leq n$. Employing eq.(5.3) for
constructing $U_+^{(n)}(R)$, from eq.(4.19) one derives
the following $n$-particle anyon hamiltonian
$$
{\widetilde H}_+^{(n)} = -{1\over 2}
\sum_{k=1}^n \left \{ \nabla_k + i\vartheta
\sum_{i=1 \atop i\neq k}^n \left [\nabla_k {\rm arg}(x_k-x_i;u)\right ]
\right \}^2 + \sum_{k=1}^n W_k \quad , \neweq
$$
which is Hermitian on ${\widetilde D}_+^n$.
This domain is invariant under the conjugation
operator $C_+$ (see eq.(2.19)),
which in the case under consideration takes the form
$$
\left [ C_+ \varphi \right ](x_1,...,x_n) =
\left \{ \exp \left [- i\vartheta \sum_{i,j =1 \atop i \not= j}^n {\rm arg }
(x_i-x_j;u) \right ] \right \}
{\overline \varphi } (x_n,...,x_1) \quad . \neweq
$$
One can also verify that the hamiltonian (5.7) commutes with $C_+$
and therefore admits self-adjoint extensions. It is known [20],
that there is actually a whole family of such extensions. The
physical meaning of this phenomenon has not been however fully clarified.

Introducing the potential
$$
A_\mu (x_k;x_1,...,{\widehat x}_k,...,x_n) =
\varepsilon_{\mu \nu } \sum_{i=1 \atop i\neq k}^n
{(x_k - x_i)^\nu \over (x_k - x_i)^2 }
\quad ,
$$
eq.(5.7) can be rewritten in the form
$$
{\widetilde H}_+^{(n)} = -{1\over 2}
\sum_{k=1}^n \left [ \nabla_k - i\vartheta
A (x_k;x_1,...,{\widehat x}_k,...,x_n) \right ]^2
+ \sum_{k=1}^n W_k  \quad . \neweq
$$
This is precisely the expression one usually
encounters in the physical literature.
Notice that the operator (5.9) is actually
well defined on the symmetric $C^\infty_0$-functions on\break
$\R^{2n}\setminus \delta $, where $\delta $ is the
subset of points $(x_1,...,x_n)$ in which two or more coordinates $x_i$
coincide.
Recently, there is some interest in the spectral problem associated with
the hamiltonian (5.9). The case $n=2$
(with $W=0$ and a harmonic potential in $|x_1 - x_2|$)
is solved exactly in the
L-M pioneering work [2]. Non-trivial three-body interactions appear for
$n\geq 3$ and the derivation of an exact solution is problematic
in that case. Some
partial results [21] and pertubative (in the parameter $\vartheta $)
computations [22] are however
available. A rigorous analysis of the spectral problem on bounded
domains in $\R^2$ has been performed in [23].

\newchapt {Multicomponent Fields}

Along the same lines, though with slight modifications, one can treat the
Fock realization of $N$-component fields satisfying eq.(1.1).
As one-particle Hilbert space we take
$$
\H = \bigoplus_{a=1}^N L^2(\R^s, d^sx) \quad . \neweq
$$
The elements of $f\in \H$ will be represented as columns with
$N$ components. The scalar product is
$$
(f,g) = \int d^sx f^{\dagger \alpha } (x) g_\alpha (x)
= \sum_{a=1}^N \int d^sx {\overline f}_\alpha (x) g_\alpha (x) \quad ,
\neweq
$$
where $\dagger $ stands for Hermitian conjugation. In this
notation an $n$-particle wave function $\varphi\in \H^n$ is a column
whose entries are
$\varphi_{\alpha_1 \cdots \alpha_n} (x_1,\dots,x_n)$.

The main ingredient in constructing $\F_R (\H )$ is the exchange factor
which for\break
$N$-component fields is an $N^2\times N^2$ matrix-valued function
on $\R^s \times \R^s $.
We assume that the entries of $R(x_1,x_2)$ are
measurable functions and impose the following additional requirements:
$$
R_{\alpha_1 \alpha_2 }^{\gamma_1 \gamma_2 }(x_1,x_2)\,
R^{\beta_1 \beta_2 }_{\gamma_1 \gamma_2 }(x_2,x_1) =
\delta_{\alpha_1 }^{\beta_1 } \, \delta_{\alpha_2 }^{\beta_2 }\quad , \neweq
$$
$$
{\overline R}_{\alpha_1 \alpha_2 }^{\beta_1 \beta_2 }(x_1,x_2) =
R^{\alpha_1 \alpha_2 }_{\beta_1 \beta_2 }(x_2,x_1)  \quad , \neweq
$$
$$
R_{\alpha_1 \alpha_2 }^{\gamma_1 \gamma_2 }(x_1,x_2)\,
R_{\gamma_2 \alpha_3 }^{\gamma_3 \beta_3 }(x_1,x_3)\,
R_{\gamma_1 \gamma_3 }^{\beta_1 \beta_2 }(x_2,x_3) =
R_{\alpha_2 \alpha_3 }^{\gamma_2 \gamma_3 }(x_2,x_3)\,
R_{\alpha_1 \gamma_2 }^{\beta_1 \gamma_1 }(x_1,x_3)\,
R_{\gamma_1 \gamma_3 }^{\beta_2 \beta_3 }(x_1,x_2)  \, .
\neweq
$$
The first two equations are the natural generalizations of eqs.(2.1,2)
and imply that $R(x_1,x_2)$ is a unitary matrix.
Eq.(6.5) is the {\it spectral} quantum Yang-Baxter equation [24], where
$\R^s$ plays the role of spectral set.
The counterpart of eq.(6.5) for $N=1$ is always satisfied and
for this reason has not been mentioned in our previous discussion.
On the
contrary, for multi-component fields eq.(6.5) is a crucial
constraint which has its origin in the associativity of the operator algebra
generated by $a_\alpha $.

The system (6.3-5) admits non-trivial solutions, but it is
a hard task to solve it in general. Since a complete
description of all solutions is presently lacking, it is instructive
to give some explicit examples. One particular solution, which can be
interpreted as a generalization of eq.(2.3), is
$$
R(x_1,x_2) = \sum_{\alpha , \beta =1 }^N
\exp [ir_{\alpha \beta }(x_1,x_2)]
E_{\alpha \beta }\otimes E_{\beta \alpha }
\quad , \neweq
$$
where $E_{\alpha \beta }$ are the Weyl matrices and
$r_{\alpha \beta }$ are real-valued functions obeying
$$
r_{\alpha \beta }(x_1,x_2) + r_{\beta \alpha }(x_2,x_1) \in 2\pi \Z \quad .
\neweq
$$
Another solution $R(x_1,x_2) = {\cal R}(x_1-x_2)$
of eqs.(6.3-5) for $x_1,x_2 \in \R^1$ is given by
$$
{\cal R}(x) = \sum_{\alpha =1}^N E_{\alpha \alpha }\otimes E_{\alpha \alpha }
+ \left ({\rm e}^{ix} - {\rm e}^{-ix}\right )
\left (q{\rm e}^{ix} - q^{-1}{\rm e}^{-ix}\right )^{-1}
\sum_{\alpha , \beta =1 \atop \alpha \not= \beta }^N
E_{\alpha \beta }\otimes E_{\beta \alpha } +
$$
$$
\left (q - q^{-1}\right )
\left (q{\rm e}^{ix} - q^{-1}{\rm e}^{-ix}\right )^{-1}
\left ( {\rm e}^{ix} \sum_{\alpha , \beta = 1\atop
\alpha > \beta }^N E_{\alpha \alpha }\otimes E_{\beta \beta } +
{\rm e}^{-ix} \sum_{\alpha , \beta = 1\atop
\alpha < \beta }^N E_{\alpha \alpha }\otimes E_{\beta \beta } \right )
\quad , \neweq
$$
where $q \in \R^1$.
The exchange matrix (6.8) stems from the quantum deformation [25,26]
of the affine Lie algebra $A^{(1)}_N$ and
has an obvious generalization to $\R^s$ by
replacing $x$ with the scalar product $x^\mu u_\mu $, $u\in \R^s$ being
an arbitrary but fixed (unit) vector. Notice also that for generic
$q\in \R^1$ the entries of ${\cal R}(x)$ are smooth functions of $x$.
Using the results of reference [26],
it is not difficult to see that the quantum deformation of the remaining
types of affine Lie algebras gives rise to solutions of eqs.(6.3-5) as
well.

Given any admissible exchange matrix, we
introduce for $n\geq 2$ the operators\break
$\{\, S_i\, :\, i=1,...,n-1\, \}$ acting on $\H^n$ according to
$$
\left [S_i\varphi \right ]_{\alpha_1 ... \alpha_n }
(x_1,...,x_i,x_{i+1},...,x_n ) =
$$
$$
\left [ R_{i\,i+1}(x_i,x_{i+1})
\right ]_{\alpha_1 ... \alpha_n }^{\beta_1 ... \beta_n }
\varphi_{\beta_1 ... \beta_n }
(x_1,...,x_{i+1},x_i,...,x_n ) \quad ,\neweq
$$
where
$$
\left [ R_{ij}(x_i,x_j)
\right ]_{\alpha_1 ... \alpha_n }^{\beta_1 ... \beta_n } =
                 \delta_{\alpha_1}^{\beta_1} \delta_{\alpha_2}^{\beta_2}
                  \cdots { \widehat {\delta_{\alpha_i}^{\beta_i}}}
                  \cdots { \widehat {\delta_{\alpha_j}^{\beta_j}}}
                  \cdots \delta_{\alpha_n}^{\beta_n}
   \,  R_{\alpha_i \alpha_j}^{\beta_i \beta_j}(x_i,x_j) \quad . \neweq
$$
Employing eqs.(6.3-5),
one can prove the $N$-component analog of proposition 1, namely
\medskip

{\bf Proposition 5}: $\{\, S_i\, :\, i=1,...,n-1\, \}$
{\it given by eq}.(6.9) {\it are bounded}
($\Vert S_i \Vert = 1$) {\it Hermitian operators on} $\H^n$
{\it and the mapping} $S\, :\, \sigma_i \mapsto S_i$
{\it is a representation of the permutation group} $\P_n$ {\it in} $\H^n$.
\medskip
At this point, the projections $P_R^{(n)}$, the $n$-particle spaces
$\H_R^{(n)}$ and the Fock space $\F_R(\H)$ for multicomponent fields are
introduced exactly as in eqs.(2.8,9,11).
The creation and annihilation operators are defined by
eq (3.1); $\a $ and $\ac $ satisfy eq.(3.2) and
act on $\F_R^0(\H)$ as follows:
$$
\left [\a \varphi \right ]_{\alpha_1\cdots \alpha_n}^{(n)}(x_1,...,x_n) =
\sqrt{n+1} \int d^sx\, f^{\dagger \alpha_0 } (x)
\varphi_{\alpha_0 \alpha_1 \cdots \alpha_n}^{(n+1)} (x, x_1,...,x_n)
\quad , \neweq
$$
$$
\left [\ac \varphi \right ]_{\alpha_1\cdots \alpha_n}^{(n)}(x_1,...,x_n) =
{1\over \sqrt{n}}
f_{\alpha_1}(x_1)\varphi_{\alpha_2\cdots \alpha_n}^{(n-1)}
(x_2,\dots ,x_n) +
$$
$$
{1\over \sqrt{n}} \sum_{k=2}^{n}\,
\left[R_{{k-1}\,k}(x_{k-1},x_k) \cdots
R_{1\,2}(x_{1},x_k)
\right]_{\alpha_1\cdots \alpha_n}^{\beta_1 \cdots \beta_n}
f_{\beta_1}(x_k)\varphi_{\beta_2\cdots \beta_n}^{(n-1)}
(x_1,\dots,{\widehat x_k},\dots,x_n)\, . \neweq
$$
The multicomponent counterpart of proposition 2 reads
\medskip

{\bf Proposition 6}: {\it The operators} $\aa$ {\it defined by eqs}.(6.11,12)
{\it satisfy} (3.10) {\it if and only if}
$$
\int d^sy \int d^sz\, g^{\dagger \alpha }(y) h^{\dagger \beta }(z)\,
R_{\alpha \beta }^{\gamma \delta }(y,z)\, g_\gamma (z) h_\delta (y) \leq 0
\neweq
$$
{\it for any} $g_\alpha ,\, h_\alpha \in B_0 (\R^s )$.
\medskip
\noindent The proof of this statement is an obvious generalization of the
argument implying the validity of proposition 2.

Introducing the operator-valued distributions $a_\alpha (x)$
and $a^{\ast \alpha }(x)$ defined by
$$
\a = \int d^sx\, f^{\dagger \alpha } (x) \, a_\alpha (x) \quad,
\quad \quad
\ac = \int d^sx\, {f}_\alpha (x) \, a^{* \alpha } (x) \quad ,
$$
from eqs.(6.11,12) one gets:
$$
a_\alpha (x_1) \, a_\beta (x_2) - R_{\beta \alpha }^{\delta \gamma }
(x_2,x_1)\, a_\gamma (x_2)\, a_\delta (x_1) = 0
\quad , \neweq
$$
$$
a^{\ast \alpha } (x_1)\, a^{\ast \beta } (x_2) -
R_{\gamma \delta }^{\alpha \beta }(x_2,x_1)\,
a^{\ast \gamma } (x_2)\, a^{\ast \delta } (x_1) = 0 \quad , \neweq
$$
$$
a_\alpha (x_1)\, a^{\ast \beta } (x_2) -
R_{\alpha \gamma }^{\beta \delta }(x_1,x_2)\, a^{\ast \gamma }(x_2)\,
a_\delta (x_1) = \delta_\alpha ^\beta \, \delta (x_1-x_2) \quad . \neweq
$$
The $R$-algebra (6.14-16) should not be confused with the so-called quon
algebra [27]. The conditions (6.3-5) imply in fact that,
apart from bosons and fermions, the set of
$R$-fields and the set of quon fields do not intersect at all.

Applying the general formalism developed in this section
to concrete exchange factors, one obtains
explicit realization of the corresponding Fock representations.
Take for example eq.(6.6) with $N=2$ and
$$
r_{11}(x_1,x_2) = r_{22}(x_1,x_2) = r(x_1,x_2) \quad ,
$$
$$
-r_{12}(x_1,x_2) = r_{21}(x_2,x_1) = r(x_1,x_2) \quad ,
$$
where $r$ satisfies (2.4). Then the algebra
(6.14-16) can be expressed entirely in terms of $R(x_1,x_2)$ given by
eq.(2.3) and as generators one can take the Wightman-type fields
$$
\phi_1 (f) = {1\over \sqrt 2 }
\left [a_1({\overline f}) + a^{\ast 2}(f)\right ]  \quad , \quad \quad
\phi_2 (f) = {i\over \sqrt 2 }
\left [a^{\ast 1}(f) - a_2 ({\overline f}) \right ] \quad ,
\neweq
$$
$$
\phi^\ast_1 (f) = {1\over \sqrt 2 }
\left [a^{\ast 1}(f) + a_2 ({\overline f})\right ]  \quad , \quad \quad
\phi^\ast_2 (f) = {-i\over \sqrt 2 }
\left [a_1({\overline f}) - a^{\ast 2} (f) \right ] \quad .
\neweq
$$
Differently from the Segal operator (3.17), the fields (6.17,18)
are by definition linear functionals in $f$.
Assuming that $R(x_1,x_2)$ is continuous for $x_1=x_2$, one derives
the exchange relations
$$
\phi_1 (x_1) \, \phi_1 (x_2) = R(x_2,x_1) \phi_1 (x_2) \, \phi_1 (x_1) \quad ,
\quad \quad
\phi_1^\ast (x_1) \, \phi_1^\ast (x_2) = R(x_2,x_1)
\phi_1^\ast (x_2) \, \phi_1^\ast (x_1)
$$
$$
\phi_1 (x_1) \, \phi_1^\ast (x_2) -
R(x_1,x_2) \phi_1^\ast (x_2) \, \phi_1 (x_1)
=\cases{
0             & if \quad $R(x,x)=1$ ,   \cr
\delta (x_1-x_2)             & if \quad $R(x,x)=-1$ , \cr}
$$
and similar equations involving also $\phi_2(x)$ and $\phi^\ast_2(x)$.
The resulting algebra generalizes the known equal-time
commutation relations of
boson ($R=1$) and fermion\break ($R=-1$) Wightman fields.

Analogously, the matrix (6.8) leads to an explicit Fock representation
for the quantized
field associated with the quantum deformation of the affine Lie
algebra $A^{(1)}_N$.

\newchapt {Conclusions}

We have analysed in the present paper some aspects of generalized
statistics in quantum field theory. In particular,
we have demonstrated that a quantum field can be associated
with any solution of the spectral Yang-Baxter
equation (6.5), satisfying the supplementary conditions (6.3,4).
The field in question admits a Fock representation with positive metric.
We have explicitly constructed the underlying Fock space
$\F_R $ and have studied its main features.
The set of all admissible exchange factors gives rise to
a whole family of fields with various statistics;
piecewise-constant (constant) $R$-matrices lead to braid (permutation) group
statistics, but the family in consideration involves also fields
corresponding to more general space-dependent exchange factors.
The physical interpretation and possible applications of the latter need
further investigation. Another point, which also deserves a more detailed
analysis is the implementation of space-time symmetries in the above
framework.

It is worth mentioning that
the explicit realization of the algebra (6.14-16) provides a new insight
into the $S$-matrix theory of integrable quantum systems (see e. g. [28]).
In fact, replacing the coordinate $x\in \R^s$ by the rapidity,
one can apply our technique for reconstructing the scattering states of
1+1 dimensional integrable models from their $S$-matrix.

Finally, the Fock representations introduced in this article suggest
some new and interesting areas of research. Among others, we have in mind
the study of concrete hamiltonian systems on $\F_R $ and the associated
quantum statistical mechanics. This subject is currently
under investigation.

\vfill\eject

\centerline {{\bf REFERENCES}}

\medskip

\item {[1]} M. G. G. Laidlaw and C. Morette DeWitt, Phys. Rev. D
{\bf 3} (1971), 1375.

\item {[2]} J. M. Leinaas and J. Myrheim, Nuovo Cimento B {\bf 37}
(1977), 1.

\item {[3]} G. A. Goldin, R. Menikoff and D. H. Sharp, J. Math. Phys.
{\bf 22} (1981), 1664.

\item {[4]} F. Wilczek, Phys. Rev. Lett. {\bf 48} (1982), 1144.

\item {[5]} Y. S. Wu, Phys. Rev. Lett. {\bf 52} (1984), 2103.

\item {[6]} J. Fr\"ohlich, {\it in} ``Non-Perturbative Quantum Field
Theory," (G. t'Hooft et al., Eds.), p. 71, Plenum Press, New York 1988;

\item {} G. Felder, J. Fr\"ohlich and G. Keller, Commun. Math. Phys.
{\bf 124} (1989), 417;

\item {} K. Fredenhagen, K-H. Rehren and B. Schroer, Commun. Math. Phys.
{\bf 125} (1989), 201.

\item {[7]} S. Deser, R. Jackiw and S. Templeton, Ann. Phys. (N.Y.)
{\bf 140} (1982), 372.

\item {[8]} J. Fr\"ohlich and P. A. Marchetti, Lett. Math. Phys.
{\bf 16} (1988), 347.

\item {[9]} R. B. Laughlin, Phys. Rev. Lett. {\bf 50} (1983), 1395;

\item {} B. I. Halperin, Phys. Rev. Lett. {\bf 52} (1984), 1583.

\item {[10]} V. Kalmeyer and R. B. Laughlin, Phys. Rev. Lett. {\bf 59}
(1987), 2095;

\item {} R. B. Laughlin, Phys. Rev. Lett. {\bf 60}
(1988), 2677;

\item {} I. E. Dzyaloshinskii, A. M. Polyakov and P. W. Wiegman, Phys.
Lett. A {\bf 127} (1988) 112.

\item {[11]} R. Jackiw and C. Rebbi, Phys. Rev. D {\bf 3} (1976), 3398;

\item {} E. Witten, Phys. Lett. B {\bf 86} (1979), 283.

\item {[12]} F. Wilczek, ``Fractional Statistics and Anyon
Superconductivity," World Scientific, Singapore 1990;

\item {} S. Forte, Rev. Mod. Phys. {\bf 64} (1992) 193.

\item {[13]} D. Arovas, J. R. Schrieffer, F. Wilczek and A. Zee, Nucl.
Phys. B {\bf 251} (1985), 117;

\item {} Y. Chen, F. Wilczek, E. Witten and B. Halperin, Int. J. Mod. Phys.
B {\bf 3} (1989), 1001.

\item {[14]} R. Jackiw and S.-Y. Pi, Phys. Rev. D {\bf 42} (1990), 3500.

\item {[15]} O. W. Greenberg and R. N. Mohapatra, Phys. Rev. D {\bf 39}
(1989), 2032.

\item {[16]} L. P. Kadanoff and H. Ceva, Phys. Rev. B {\bf 3} (1971), 3918;

\item {} G. 't Hooft, Nucl. Phys. B {\bf 138} (1978), 1.

\item {[17]} D. Hasenfratz, Phys. Lett. B {\bf 85} (1979), 338.

\item {[18]} M. Reed and B. Simon, ``Methods in Modern Mathematical Physics
I: Functional Analysis," Academic Press, New York, 1972.

\item {} M. Reed and B. Simon, ``Methods in Modern Mathematical Physics
II: Fourier Analysis, Self-Adjointness," Academic Press, New York, 1975.

\item {[19]} M. Mintchev and M. Rossi, Phys. Lett. B {\bf 271}
(1991), 187;

\item {} A. Liguori, M. Mintchev and M. Rossi, Phys. Lett. B {\bf 303}
(1993), 38;

\item {} A. Liguori, M. Mintchev and M. Rossi, Phys. Lett. B {\bf 305}
(1993), 52.

\item {[20]} P. de Sousa Gerbert and R. Jackiw,  Commun. Math. Phys.
{\bf 124} (1989), 229;

\item {} B. S. Kay and U. M. Studer, Commun. Math. Phys. {\bf 139}
(1991), 103.

\item {[21]} C. Chou, Phys. Rev. D {\bf 44} (1991), 2533; Erratum: Phys.
Rev. D {\bf 45} (1992), 1433;

\item {} G. V. Dunne, A. Lerda, S. Sciuto and C. A. Trugenberger, Nucl.
Phys. B {\bf 370} (1992), 601.

\item {[22]} J. McCabe and S. Ouvri, Phys. Lett. B {\bf 260} (1991), 113.

\item {[23]} G. A. Baker, G. S. Canright, S. B. Mulay and C. Sundberg,
Commun. Math. Phys. {\bf 153} (1993), 277.

\item {[24]} C. N. Yang, Phys. Rev. Lett. {\bf 19} (1967), 1312;

\item {} R. J. Baxter, ``Exactly Solved Models in Statistical Mechanics,"
Academic Press, New York, 1982.

\item {[25]} V. G. Drinfeld, {\it in} "Proceedings of the International
Congress of Mathematics", (A. M. Gleason, Ed.), p. 798, American
Mathematical Society, Providence 1987.

\item {[26]} M. Jimbo, Commun. Math. Phys. {\bf 102} (1986), 537.

\item {[27]} O. W. Greenberg, Phys. Rev. D {\bf 43} (1991), 4111.

\item {[28]} F. A. Smirnov, ``Form Factors in Completely Integrable Models
of Quantum Field Theory," World Scientific, Singapore 1992.


\vfill\eject
\bye